\documentclass[12pt]{article}

\usepackage[]{graphicx,float,latexsym,times}
\usepackage{amsfonts,amstext,amsmath,amssymb,amsthm, bm,mathrsfs}
\usepackage{multirow}
\usepackage{xcolor}
\usepackage{setspace}
\usepackage{caption}
\usepackage{subcaption}

\long\def\symbolfootnote[#1]#2{\begingroup
\def\thefootnote{\fnsymbol{footnote}}\footnote[#1]{#2}\endgroup}

\usepackage{url} 
\usepackage{bbm}

\usepackage{titlesec} 
\titleformat{\section}{\large\bfseries}{\thesection.}{.5em}{}
\titlespacing*{\section}{0pt}{*3}{*2}
\titleformat{\subsection}{\normalfont\bfseries}{\thesubsection.}{.5em}{}
\titlespacing*{\subsection} {0pt}{*3}{*2}
\titleformat{\subsubsection}{\normalfont\bfseries}{\thesubsubsection.}{.5em}{}
\titlespacing*{\subsubsection} {0pt}{*3}{*2}

\theoremstyle{plain} 
\newtheorem{theorem}{Theorem}[section]
\newtheorem{lemma}{Lemma}[section]

\theoremstyle{definition} 

\newtheorem{remark}{Remark}[section]

\numberwithin{equation}{section} 

\usepackage[authoryear]{natbib}

\newcommand{\calH}{{\mathcal H}}
\newcommand{\calC}{\mathcal C}
\newcommand{\calI}{{\mathcal I}}
\newcommand{\calN}{{\mathcal N}}

\newcommand{\bbR}{{\mathbb R}}

\newcommand{\bbE}{{\mathbb E}}
\newcommand{\bfu}{{\mathbf u}}
\newcommand{\bfs}{{\mathbf s}}

\newcommand{\Var}{{\rm Var}}
\newcommand{\Cov}{{\rm Cov}}

\begin{document}

       \title{\textbf{\Large Online Score Statistics for Detecting Clustered Change in Network Point Processes}}
        \date{}
        \maketitle
       \author{
\begin{center}
\vskip -1cm
\textbf{\large Rui Zhang*, Haoyun Wang*, Yao Xie}

School of Industrial and Systems Engineering (ISyE), \\Georgia Institute of Technology, Atlanta, Georgia, USA.
\end{center}
}
\symbolfootnote[0]{\normalsize Address correspondence to Yao Xie,
School of Industrial and Systems Engineering, Georgia Institute of Technology, Atlanta, Georgia, 30332, USA; E-mail: yao.xie@isye.gatech.edu. \\ * Authors have equal contribution.}

{\small \noindent\textbf{Abstract:} 
We consider online monitoring of the network event data to detect local changes in a cluster when the affected data stream distribution shifts from one point process to another with different parameters. Specifically, we are interested in detecting a change point that causes a shift of the underlying data distribution that follows a multivariate Hawkes process with exponential decay temporal kernel, whereby the Hawkes process is considered to account for spatio-temporal correlation between observations. The proposed detection procedure is based on scan score statistics. We derive the asymptotic distribution of the statistic, which enables the self-normalizing property and facilitates the approximation of the instantaneous false alarm probability and the average run length. When detecting a change in the Hawkes process with non-vanishing self-excitation, the procedure does not require estimating the post-change network parameter while assuming the temporal decay parameter, which enjoys computational efficiency. We further present an efficient procedure to accurately determine the false discovery rate via importance sampling, as validated by numerical examples. Using simulated and real stock exchange data, we show the effectiveness of the proposed method in detecting change while enjoying computational efficiency.}

\section{INTRODUCTION}
    
Network Hawkes point processes recently became a popular model for sequential events data over networks, a widely encountered data type in modern applications. Such data usually capture the temporal and spatial information of the events, i.e., a sequence of event times, the corresponding event location, and additional information. Multivariate Hawkes process can model the influence of the previous events on the subsequent events, for instance, exciting the subsequent events or causing them more likely to happen. Hawkes processes have been widely used in many areas such as finance \citep{hawkes2018hawkes}, social media \citep{rizoiu2017hawkes}, epidemiology \citep{rizoiu2018sir}, seismology \citep{ogata1998space}.
    
Change-point detection is a fundamental problem in statistics; aiming to detect a transition in the distribution of the sequential data, which often represents a state transition (see, e.g., \cite{Moustakides2008,Poor2008,XieSiegmund2013,VVV2014,Tartakovsky2019, Xie2021survey}). For example, 
in water quality monitoring, a change-point can be due to a water contamination event \citep{chen2020S3T}; in public health, they may represent a disease outbreak \citep{chris2010social}. The goal is to develop a procedure that can raise the alarm as soon as possible after the change-point while controlling the false-alarm rate.

Detecting change-points for Hawkes processes is an important problem for monitoring large-scale networks using discrete events data. Online detection of change-points detection in Hawkes processes is considered challenging due to the asynchronous nature of discrete events and temporal dependence, which is far from the traditional i.i.d. setting considered in change-point detection literature. In \cite{wang2021sequential}, the authors propose an adapted version of the CUSUM procedure to account for the dynamic behavior assuming known post-change parameters. When post-change parameters are unknown, a classic method is the generalized likelihood ratio (GLR) procedure, and \cite{li2017detecting} use the Expectation-Maximization type of algorithm to estimate the post-change parameters of Hawkes processes and then compute the generalized likelihood ratios. This method needs large memory and computation time to obtain the maximum likelihood estimates at each time instance. An alternative to GLR is the score test \citep{Rao2005}, which does not require computing the MLE. The score test is well studied and widely applied. In the univariate case, the score test is the most powerful test for small deviation from the null hypothesis. For multi-stream network data, a strategy is needed to combine the high dimensional statistic -- \cite{xie2021optimality} proposed a graph scanning statistics by computing the statistics for subgraphs which is helpful to detect and identify the local changes. Besides online change-point detection, recent work by \cite{wang2020detecting} also developed methods and established theoretical guarantees for offline change-point detection for Hawkes processes. 

In this paper, we present a graph scan score statistic for detecting local changes that happen as a cluster in the network when observing sequential discrete event data that can be modeled using point processes. The change causes an unknown shift in the underlying parameter of the Hawkes process over a sub-network. We assume the parameters of the pre-change distributions are known since typically abundant ``normal'' and ``in-control'' data can be used to estimate the pre-change parameters with good precision. We assume the post-change parameters are unknown since they are typically due to an unexpected anomaly. This motivates us to consider the score statistic, which detects a departure from the ``normal'' data without having to estimate the post-change parameters. We present the asymptotic distribution of the score statistic, which enables us to develop the self-normalizing scan statistic over pre-defined candidate scanning clusters. This also leads to an accurate approximation of the instantaneous false alarm probability, the false alarm rate, the average run length and an efficient procedure to accurately determine the false discovery rate via importance sampling, as validated by numerical examples. The good performance of our procedures compared with the benchmarks is tested with numerical experiments with simulated and real stock exchange data.
    
The rest of our paper is organized as follows. Section 2 provides the background knowledge of the multivariate Hawkes process. Section 3 presents the definition of our problem. Section 4 proposes our detection procedure and includes the analysis of our scan score statistics. In section 5, there are experiments of a simulation study and real-world data application. Section 6 concludes our paper.

\section{PRELIMINARIES}

A multivariate Hawkes process is a self-exciting process over a network. Let $M$ denote the number of nodes in the network and $[M]$ denote $\{1,\dots, M\}$. The data is of the form $\{(u_1, t_1), (u_2, t_2),\dots\}$, where $u_i\in[M]$ denotes the location of the $i$th event and $t_i\in\bbR^+$ denotes the time of the $i$th event. A multivariate Hawkes Process is actually a special case of spatio-temporal counting process \citep{rathbun1996asymptotic}. Let $\calH_t$ denote the history before time $t$, i.e. the $\sigma$-algebras of events before time $t$; $\{\calH_t\}_{t\ge 0}$ is a filtration, an increasing sequence of $\sigma$-algebras. Let $N_m(t)$ denote the number of events on $i$th node up to time $t$, i.e., a counting process,
    \[N_m(t) = \sum_{t_i\leq t} \mathbb I(t_i\leq t, u_i = m),\]
    where $\mathbb I$ denotes the indicator variable. 
    Then, a multivariate Hawkes process can be determined by the following conditional intensity function \citep{reinhart2018Arev}:
    \begin{equation}
        \lambda_m(t) = \lim_{s\rightarrow 0}\frac{\mathbb P\{N_m(t + s)>0|\calH_t\}}{s}.
    \end{equation}
    For a multivariate Hawkes process, the conditional intensity function takes the form:
    \begin{equation}
    \label{lambda}
        \lambda_m(t) = \mu_m + \sum_{i\in[M]}\int_{0}^t g_{i,m}(t-s) N_i(ds).
    \end{equation}
    Here $\mu_m$ is the base intensity and $g_{i,j}(t)$ is the kernel function that characterized the influence of the previous events. Specifically, we assume a commonly used exponential kernel, i.e.,
    \begin{equation}
    \label{kernel}
        g_{i,j}(t) = \alpha_{i,j}e^{-\beta t},
    \end{equation}
    where $\beta>0$ is a parameter that controls the decay rate. 
    Let $\bm \mu = (\mu_1,\dots, \mu_M)$ and $\mathbf A\in\bbR^{M\times M}$, of which the $(i,j)$th entry is $\alpha_{i,j} \geq 0$. A multivariate Hawkes process with exponential kernel is parametrized by the base intensity $\bm\mu$, influence matrix $\mathbf A$ and decay rate $\beta$. Given all the events in time window $[0, T]$, the log likelihood function is given by:
    \begin{equation}\label{eq:llh}
    \begin{split}
        \ell_T(\mathbf A)  =& \sum_{k=1}^K\log\Big(\mu_{u_k}+\sum_{t_i<t_k}\alpha_{u_i,u_k}
        e^{-\beta(t_k-t_i)}\Big) - \sum_{m=1}^M\mu_mT\\
                     &+\frac{1}{\beta}\sum_{m=1}^M\sum_{k=1}^K\alpha_{u_k,m}[e^{-\beta(T-t_k)}-1],
    \end{split}
    \end{equation}
    where $K$ denote the number of events before time $T$. Note that when $\mathbf A = \mathbf 0$, the process becomes a multivariate Poisson process.

\section{PROBLEM SETUP}
 
Consider a network with $M$ nodes; we can observe a sequence of events on each node over time. There may exist a change-point in time $\tau^*>0$ if the following applies. Before time $\tau^*$, the events of the network follow a multivariate Hawkes process withparameters $\bm\mu$, $\mathbf A_0$, and $\beta$.  After time $\tau^*$, the events of the network follow a multivariate Hawkes of which the influence matrix change from $\mathbf A_0$ to $\mathbf A_1$. The pre-change and post-change multivariate Hawkes processes are assumed to be stationary, i.e., $\|\mathbf A_0\|< 1$ and $\|\mathbf A_1\|<1$ where $\|\cdot\|$ represents the spectral norm. To detect whether a change-point $\tau^*$ exists in the given data, we consider the following hypothesis test:
    \begin{equation}
        \label{HT}
        \begin{split}
            H_0: &\lambda_m(t) = \mu_m + \sum_{t_i\leq t} \alpha_{u_i, m,0}e^{-\beta(t-t_i)}, m\in [M], t\ge 0;\\
            H_1:&\lambda_m(t) = \mu_m + \sum_{t_i\leq \tau^*} \alpha_{u_i, m,0}e^{-\beta(t-t_i)}, m\in [M], 0\leq t\leq \tau^*;\\
            &\lambda_m(t) = \mu_m + \sum_{\tau^*\leq t_i\leq t} \alpha_{u_i, m,1}e^{-\beta(t-t_i)}, m\in [M], t> \tau^*;
        \end{split}
    \end{equation}
where $\lambda_m(t)$ is the true conditional intensity of node $m$ at time $t$, $\alpha_{i,j,0}$ and $\alpha_{i,j,1}$ are the $(i,j)$th entry of $\mathbf A_0$ and $\mathbf A_1$, respectively. In particular, we refer to $\mathbf A_0$ and $\mathbf A_1$ as the network parameters as they describe the interactions (influences) between nodes in the network.

\section{SCAN SCORE STATISTICS DETECTION PROCEDURE}
    
To perform the sequential hypothesis test (\ref{HT}), we proposed a detection procedure base on scan score statistics. A score statistics corresponds to the first-order derivative of the log-likelihood function. In a multivariate Hawkes network, we are interested in the influence between multiple pairs of nodes (i.e., the entries in the influence matrix $\mathbf A$). For each pair, we can derive the score statistics which leads to a multi-dimensional vector of score statistics. Based on that, we use a scanning strategy to compute our test statistics, similar to \cite{chen2020S3T, he2018sequential}. Specifically, we divide the network into several clusters, compute the score statistics in each cluster at each time $t$, and then obtain a detection statistic for each cluster by summing up the standardized score statistics in the corresponding cluster. Finally, we take the maximum over all the clusters to form the scan score statistics at time $t$ for the entire network. More details will be discussed in this section. 

\subsection{Score Statistics}

Since the change in hypothesis test (\ref{HT}) is caused by the change of influence matrix, we define the following score statistics, given data up to time $t$, with respect to $\alpha_{p,q}$:
    \begin{equation}
    \label{score1}
        S^{(p,q)}_T(\mathbf A) \triangleq \frac{\partial \ell_T(\mathbf A)}{\partial \alpha_{p,q}}.
    \end{equation} 
    Moreover, define $S_T(\mathbf A)$ is the vector of all elements in  $\{S^{(p,q)}_T(
    \mathbf A); p,q\in[M]\}$. According to Theorem 1 in \cite{rathbun1996asymptotic}, we have the following corollary.
    \begin{lemma}\label{lemma1}
        Under the assumptions in \cite{rathbun1996asymptotic}, assume the influence matrix of the multivariate Hawkes Process is $\mathbf A$. The score function $S_T(\mathbf A)$ satisfies that, 
        \[T^{-\frac{1}{2}} S_T(\mathbf A)\stackrel{D}{\rightarrow}\calN(0, \calI(\mathbf A)),\] where $\calI(\mathbf A)$ is the Fisher information. 
    \end{lemma} 
    
    \subsubsection{Theoretical Characterization of Fisher Information $\calI(\mathbf 0)$}
    
    When $\mathbf A = \mathbf 0$, it is possible to compute $\calI(\mathbf 0)$ as shown in the following theorem. For simplicity, let's define $\calC(i,t)$ as the set of events at node $i$ before time $t$, i.e. $\calC(i,t) = \{k: t_k<t, u_k = i\}$.
    \begin{theorem}
    \label{thm1}
     Assume the the conditional intensity function has the form as in eq.(\ref{lambda}) and the kernel function is exponential as in eq.(\ref{kernel}). According to eq.(\ref{score1}),
     \begin{equation}
             \begin{split}
        S_T^{(p,q)}(\mathbf A)  =& \frac{\partial \ell_T(\mathbf A)}{\partial\alpha_{p,q}}\\
                        =& \sum_{k\in\calC(q,T)} \frac{\sum_{i\in\calC(p,t_k)}e^{-\beta(t_k-t_i)}}
        {\mu_q+\sum_{t_i<t_k}\alpha_{u_i,q}e^{-\beta(t_k-t_i)}}+\frac{1}{\beta}\sum_{k\in\calC(p,T)}
        [e^{-\beta(T-t_k)}-1].
        \end{split}
     \end{equation}
     Moreover, when $\mathbf A = \mathbf 0$, as $T\rightarrow\infty$, the non-zero elements in the limit of variance (i.e. $\calI(\mathbf A)$ in Lemma \ref{lemma1}) are as following:
        \begin{equation}
            \begin{split}
        &\Var[T^{-\frac{1}{2}}S_T^{(q,q)}(\mathbf 0)] \rightarrow \frac{1}{2\beta} + \frac{\mu_q}{\beta^2},\\
        &\Var[T^{-\frac{1}{2}}S_T^{(p,q)}(\mathbf 0)] \rightarrow \frac{\mu_p}{\mu_q}(\frac{1}{2\beta} + \frac{\mu_p}{\beta^2}),\\
        &\Cov[T^{-\frac{1}{2}}S_T^{(p,q)}(\mathbf 0), T^{-\frac{1}{2}}S_T^{(p',q)}(\mathbf 0)] \rightarrow \frac{\mu_p\mu_{p'}}{\mu_q\beta^2}.
            \end{split}
        \end{equation}
    \end{theorem}
    %
    
    \subsubsection{Estimation of $\calI(\mathbf A)$}
    
    When $\mathbf A\neq\mathbf 0$, it is difficult to compute the variance
    theoretically. According to \cite{rathbun1996asymptotic}, we have the following approximation of $\calI (\mathbf A)$.
    \begin{theorem}\label{thm2}
    With the same assumption as in Theorem \ref{thm1}. We have the following estimation of fisher information. Let
    \begin{equation}\label{eq:estI}
        \hat\calI_{T}(\mathbf A)_{(i,j),(p,q)} = \begin{cases}
            0 \,\,\, &{\rm if}\,\,\,j\neq q,\\
            \sum_{k\in\calC(q,T)}\frac{(\sum_{k\in\calC(i,t)}e^{-\beta(t_k-t_i)})(\sum_{k\in\calC(p,t)}e^{-\beta(t_k-t_i)})}
            {\big(\mu_q+\sum_{t_i<t_k}\alpha_{u_i,q}e^{-\beta(t_k-t_i)}\big)^2} &{\rm if}\,\,\, j=q.
        \end{cases}
    \end{equation}
    We have $\frac{1}{T}\hat\calI_T(\mathbf A)\rightarrow \calI(\mathbf A)$, i.e. $\forall i, j,p,q$, 
    \begin{equation}
        \frac{1}{T}\hat\calI_T(\mathbf A)_{(i,j), (p,q)}\rightarrow \calI(\mathbf A)_{(i,j),(p,q)},
    \end{equation}
    where $\calI(\mathbf A)_{(i,j),(p,q)}$ is the asymptotic (co)variance of $T^{-1/2}S_T^{(i,j)}(\mathbf A)$ and $T^{-1/2}S_T^{(p,q)}(\mathbf A)$.
    \end{theorem}

    \subsection{Scan Score Statistics}
    
    To combine all the score statistics and complete the detection procedure, we compute the scan statistics based on given clusters. A cluster is a directed subgraph, with the set of nodes $V_i$ and the set of edges $E_i$, $i=1,2,\dots,L$ where $L$ is the number of clusters. In practice, to reduce the computation cost, we only compute the score statistics given data in a time window of length $w$, and update the statistics every $\delta$ time units, where $\delta\leq w$. Specifically, at time $t$ which is a multiple of $\delta$, for the $i$th cluster, we compute all the interested score statistics at the pre-change parameter $\mathbf A_0$ with data in $[t-w, t]$ and have a vector of score statistics denoted as $S^{(i)}_{t,w}(\mathbf A_0) = (S_t^{(p,q)}(\mathbf A_0) - S_{t-w}^{(p,q)}(\mathbf A_0))_{(p,q)\in E_i}$. Let $R_i$ and $\mathcal I^{(i)}(\mathbf A_0)$ denote the dimension and Fisher information corresponds to edges in $E_i$ respectively. Then the detection statistics for cluster $i$ at time $\tau$, with window length $w$ is:
    \begin{equation}
    \label{eq-scanStat}
        \Gamma_{t, w}^{(i)} =  (wR_i)^{-1/2}\mathbf 1^\top \mathcal I^{(i)}(\mathbf A_0)^{-1/2}S_{t, w}^{(i)}(\mathbf A_0) \sim \mathcal N(0, 1).
    \end{equation}
    Note that the scan statistic has a ``self-normalizing'' property in that their asymptotic distributions are standard normal for multiple candidate clusters, which will facilitate the false alarm control by choosing a threshold. For example if we are interested in detecting a shift from a Poisson process, all we need is to evaluate $\calI(\mathbf A_0)$, which only requires the pre-change parameter without having to estimate the post-change parameter $\mathbf A_1$. The estimation of $\mathbf A_1$ can be difficult to perform online given limited post-change observations, since we would like to detect the change quickly. 
    
    Then at each time $t$, we compute the scan score statistics over the candidate clusters:
    \[\Gamma_{t} = \max_{1\leq i\leq L} |\Gamma_{t, w}^{(i)}|.\] 
    Given a threshold $b > 0$, we stop our procedure and raise an alarm to detect a local change-point using the following rule:
    \begin{equation}\label{eq:Stop-time}
        T_b = \inf\{t: \Gamma_t>b\}.
    \end{equation}

    In the following, we discuss the false alarm probability at any instant $t$ in Section \ref{false-alarm-rate}, then provide the performance analysis of $T_b$ in Section \ref{sec:performance_analysis}. Here the choice of the threshold $b$ controls the tradeoff between the false alarm rate and average run length versus the detection delay. The proposed method can also be used to localize the change once an alarm is raised, and we briefly discuss the false discovery rate in \ref{sec-fdr}.
    
    \subsubsection{Instantaneous False Alarm Probability of Scan Statistics at a Given $t$} \label{false-alarm-rate}

\begin{figure}[H]
        \centering
        \begin{minipage}{0.4\textwidth}
        \centering
        \includegraphics[width =\textwidth]{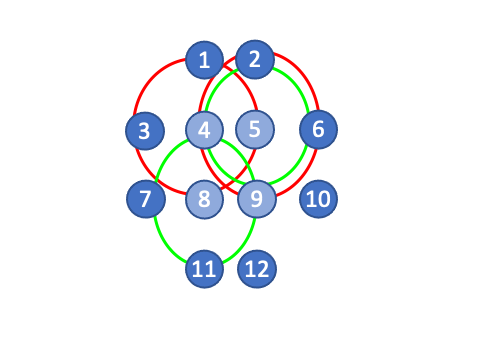}
        \subcaption[]{}
        \label{fig:exaAll}
        \end{minipage}
        \begin{minipage}{0.4\textwidth}
        \centering
        \includegraphics[width =\textwidth]{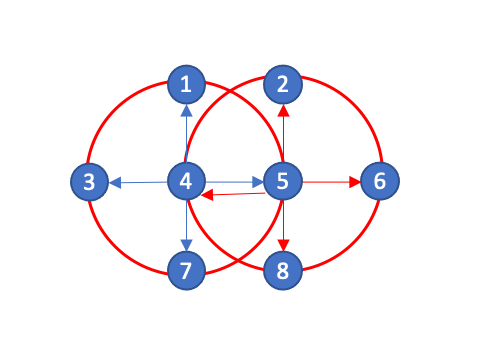}
        \subcaption[]{}
        \label{fig:exaI}
        \end{minipage}
        \begin{minipage}{0.4\textwidth}
        \centering
        \includegraphics[width= \textwidth]{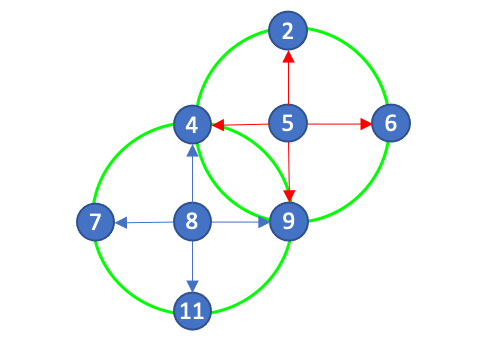}
        \subcaption[]{}
        \label{fig:exaII}
        \end{minipage}
        \caption{In this example, there are 4 clusters, and each cluster includes 5 locations. The 4 clusters are $(1,3,4,5,8)$, $(2,4,5,6,9)$, $(4,7,8,9,11)$ and $(5,8,9,10,12)$. The light blue nodes are the centers of each cluster and in each cluster we consider the 4 directions from the center to the around as shown in \textbf{\emph{(b)}} and \textbf{\emph{(c)}}. $S^{(i)}_{t,w}\sim \mathcal N(\mathbf 0_4, w(1/(2\beta)+\mu/\beta^2)\mathbf I_4)$ and $\Gamma_{t,w}\sim\mathcal N(0,\Sigma)$. For the case in \textbf{\emph{(b)}}, the $\Sigma_{ij}$ corresponding to the covariance between $\Gamma^{(i)}_{t,w}$ and $\Gamma^{(j)}_{t,w}$ equals to 0. For the case in \textbf{\emph{(c)}}, $\Sigma_{ij} = \sigma^2 \triangleq \mu/(\beta+2\mu)$. Therefore, $\Sigma = ((1,0, 0, \sigma^2)^\top,(0,1,\sigma^2, 0)^\top,(0, \sigma^2,1, 0)^\top,(\sigma^2, 0,0,1)^\top)$.}
        \label{fig:exaSig}
    \end{figure}

    According to Lemma \ref{lemma1}, the score $T^{-1/2}S_T(\mathbf A_0)$ converge in distribution to $\mathcal N(0,\mathcal I(\mathbf A_0))$ as $T\to \infty$. Since the Hawkes process is stationary, $w^{-1/2}(S_{w\tau}(\mathbf A_0) - S_{w(\tau-1)}(\mathbf A_0))$ also converge to the same distribution as $w\to\infty$. The statistics $\Gamma_{t,w}^{(i)}$ are linear combinations of $w^{-1/2}(S_{t}(\mathbf A_0) - S_{t-w}(\mathbf A_0))$, then after scaling the time by $w^{-1}$, the process $(\Gamma_{w\tau,w}^{(i)})_{\tau\geq 0,1\leq i\leq L}$ converge pointwise to a Gaussian process as $w\to\infty$, and the covariance can be characterized by  
    \begin{equation}\label{eq:gammaCov}
        \Cov(\Gamma^{(i)}_{w\tau,w}, \Gamma^{(j)}_{w(\tau+\epsilon), w}) = (1-\epsilon)^{+} \Cov(\Gamma^{(i)}_{1,1}, \Gamma^{(j)}_{1,1}),\text{ for any $1\leq i\leq j\leq L$, for any $\epsilon\geq 0,$}
    \end{equation}
    where the covariance between $\Gamma_{1,1}^{(i)}$ and $\Gamma_{1,1}^{(j)}$ can be found in close form when $\mathbf A_0 = \mathbf 0$ using Theorem \ref{thm1} and estimated using historical data by Theorem \ref{thm2}. At time $t > 0$, we want to control the instantaneous false alarm probability
    \begin{eqnarray}
    \label{eq:maxNorm}
        \mathbb P(\Gamma_t>b) &=& \mathbb P\big( \max_{1\leq i\leq L} |\Gamma^{(i)}_{t,w}| \ge b\big)
        = \mathbb P\big( \bigcup_{i=1}^L |\Gamma^{(i)}_{t,w}| \ge b\big)\nonumber\\
        &=&\mathbb P\big( \bigcup_{i=1}^L \{\Gamma^{(i)}_{t,w} \ge b\}\bigcup_{i=1}^L \{\Gamma^{(i)}_{t,w} \leq -b\}\big)\nonumber\\
         &=&\mathbb P\big(  \{\max_{1\leq i\leq L} \Gamma^{(i)}_{t,w} \ge b\} \cup \{\min_{1\leq i\leq L} \Gamma^{(i)}_{t,w} \leq -b\})\nonumber\\
         &\leq&  2\mathbb P\big( \max_{1\leq i\leq L} \Gamma^{(i)}_{t,w} \ge b\big).
    \end{eqnarray}
   Let $\boldsymbol\Gamma_{t,w}$ denote the vector of $\Gamma^{(i)}_{t,w}$s. To control the upper bound of the instantaneous false alarm probability, we compute \eqref{eq:maxNorm} with the technique in \cite{botev2015tail}:
    \begin{align}\label{eq-tailProb}
        \mathbb P\big( \max_{1\leq i\leq L} \Gamma^{(i)}_{t,w} \ge b\big) &= \mathbb P\big(\bigcup_{i=1}^{L} \{\Gamma^{(i)}_{t,w}\ge b, \Gamma^{(i)}_{t,w}\ge\Gamma^{(j)}_{t,w}, j \neq i\}\big)\nonumber\\
        &= \sum_{i=1}^L\mathbb P \big( \Gamma^{(i)}_{t,w}\ge b, \Gamma^{(i)}_{t,w}\ge\Gamma^{(j)}_{t,w}, j \neq i\}\big)\nonumber\\
        &=\sum_{i=1}^L\mathbb P\big(\mathbf B\mathbf P_i\boldsymbol\Gamma_{t,w} \ge \mathbf b\big),
    \end{align}
    where $\mathbf P_i$ is the permutation matrix interchanging first entry and the $i$th entry, and
    \begin{eqnarray}
        \mathbf B = \left[\begin{array}{ccccc}
             1&0&\cdots&\cdots&0  \\
             1&-1&0&\cdots&0  \\
             1&0&-1&\ddots&\vdots  \\
             \vdots&\vdots&\ddots&\ddots&0  \\
             1&0&\cdots&0&-1  \\
        \end{array}\right],\,\,\, \mathbf b = \left(\begin{array}{c}
              b \\
                0\\
                \vdots\\
                0
        \end{array}\right).
    \end{eqnarray}
    Here $\boldsymbol\Gamma_{t,w}$ follows a Gaussian distribution where the covariance $\Sigma$ can be computed with (\ref{eq:gammaCov}) according to the network topology and the score statistics in the clusters. In \cite{botev2015tail}, the authors provide an importance sampling algorithm to estimate \eqref{eq-tailProb}. Figure \ref{fig:exaSig} is an example of the cluster structure and the corresponding $\Sigma$ when $\mathbf A_0 = \mathbf 0$.
 
    \begin{remark}
    Since our scan statistics are standardized, determining the threshold $b$ with the method in eq.(\ref{eq-tailProb}) does not depend on the window length $w$. However, with a larger $w$ the Gaussian process approximation would be better. In Table \ref{tab:FARAppr}, we can see, as window length increases, the instantaneous false alarm probability would be better controlled.
    \end{remark}
    \begin{table}[H]
        \begin{center}
        \caption{Accuracy of approximation of instantaneous false alarm probability through \eqref{eq-tailProb}.}
        \begin{tabular}{|c|c|c|c|}
        \hline
             $w$&$b$&$2\mathbb P(\max_i \Gamma^{(i)}_{t,w}>b)$&$\mathbb {\hat P}(\Gamma_{t}>b)$  \\\hline
             50& 3 & $0.01$ & 0.0174\\\hline
             100& 3 & $0.01$ & 0.0146\\\hline
             200& 3 & $0.01$ & 0.0114\\\hline
             50& 2.8 & $0.02$ & 0.0282\\\hline
             100& 2.8 & $0.02$ & 0.0226\\\hline
             200& 2.8 & $0.02$ & 0.0210\\\hline
        \end{tabular}
        \label{tab:FARAppr}
        \end{center}
    \end{table}

    \subsubsection{Performance metrics of the stopping time $T_b$}
    \label{sec:performance_analysis}
    In this part, we provide an upper bound of the false alarm rate (FAR) and average run length (ARL) based on the Gaussian process approximation and the analysis of the instantaneous false alarm probability. Then we discuss the detection delay and the choice of the window length $w$ and update interval $\delta$.
    
    The FAR is the conditional probability that the procedure will stop at the next update, given that there has been no false alarm yet.
    $$
    \mbox{FAR} = \sup_n(T_b = (n+1)\delta|T_b>n\delta).
    $$
    We have the following result on the FAR.
    \begin{theorem}\label{TheoremFAR}
    \label{thm4}
    As $b\to \infty$, if $w/\delta$ is upper bounded by some constant $C$,
    $$
    \frac{\mbox{\textup{FAR}}}{\mathbb P(\Gamma_t>b)} \leq 1+o(1).
    $$
    \end{theorem}
    
    The ARL of $T_b$ is $\mathbb E[T_b]$, the expected stopping time under $H_0$. To evaluate the ARL, we are going to show that $T_b$ is approximately exponential distribution with some parameter $\lambda_0$. The analysis is similar to \cite{yakir2009multi}. Let $f(b) = be^{b^2/2}$, such that $1/f(b)$ is in the same order as the instantaneous false alarm probability. For any $x>0$ and interval $[0, x f(b)\delta]$, we decompose it into $k$ sub-intervals with length $m\delta$, i.e $xf(b)=km$.  For simplicity, we assume $k$ and $m$ are integers. 
    
    Let indicator $X_j$ denotes $\mathbb I\{\max_{(j-1)m<n\leq jm} \Gamma_{n\delta}>b\}$, and define $W = \sum_{j=1}^k X_j$, then we have \[\{W = 0\} = \{T_b>xf(b)\delta\}.\] To prove that $T_b$ is approximately exponential, it is same to prove $W$ is approximately Poisson distributed.  We herein apply the result from \cite{arratia1989two}. According to the Theorem I in \cite{arratia1989two}, we establish the following theorem.
    \begin{theorem}\label{thm3}
    Let $T_b$ be the stopping time defined in eq.(\ref{eq:Stop-time}), $X_j$ be the indicator defined above and $W$ be the sum of the indicators. With $w/\delta\ll m\ll f(b)$, for any fixed $x\geq 0$,
        \begin{equation}
            \mathbb \lim_{b\to \infty } |\mathbb P(T_b>xf(b)\delta)-e^{-\mathbb E W}| = 0.
        \end{equation}
    \end{theorem}
    The theorem above can be used to obtain an approximation of ARL. According to the construction of $W$, we have
    \begin{equation*}
    \begin{split}
    \mathbb E W & = k\mathbb P(X_j =1) \\
    & = xf(b)\mathbb P\{\max_{0<n\leq m} \Gamma_{n\delta}>b\}/m \\
    & \leq xf(b)\mathbb P\{\Gamma_{t}>b\}.
    \end{split}
    \end{equation*}
     By Theorem \ref{thm3}, $\mathbb E_{\infty}(T_b)\approx \lambda_0^{-1}\delta$ and 
    \begin{eqnarray}
        \lambda_0&\leq&\mathbb P\{\max_{0<n\leq m} \Gamma_{n\delta}>b\}/m \label{eq:est1}\\
        &\leq&\mathbb P(\Gamma_{t}>b).\label{eq:est2}
    \end{eqnarray}
    Therefore, we can use our instantaneous false alarm probability approximation in Section \ref{false-alarm-rate} to approximate \eqref{eq:est1}. We can numerically verify that this is a reasonably accurate approximation, in Section \ref{exp:ARL}.

    \begin{remark}
    To evaluate the performance of our scanning statistics, we also need the expected detection delay (EDD). Since we are using a Shewhart chart type of detecting procedure, the EDD varies with the window length. In practice, the window length $w$ can be chosen by considering the smallest change we want to detect on each cluster. Then $w$ will be the smallest window length that has enough power to detect the change successfully, i.e., under the post-change scenario, $ |\mathbb E(\Gamma_{t,w}^{(i)})|\geq b$ if the change happens on the $i$-th cluster. 
    \end{remark}
    
    \begin{remark}
    The performance of our scanning statistic also depends on the update interval $\delta$, where a smaller $\delta$ results in both a smaller ARL and EDD. However, there seems to be little point in choosing $\delta \ll w$ to frequently check for a potential change-point while sacrificing computation efficiency because we expect the change is not too large and cannot be reflected immediately in the detecting statistic $\Gamma_t$.
    \end{remark}

    \section{EXPERIMENTS}
    \subsection{Simulated Result of ARL and EDD}\label{exp:ARL}
    In this experiment, the network is set up as shown in Figure \ref{fig:exaSig}. The event in each node follows a Poisson process with $\mu = 1$, and we set the $\beta = 1$. The window length is set to be $200$, and the statistics are computed for each $\delta = 10$ time unit. In Table \ref{tbl:ARL}, we show the estimated ARL from simulation with the threshold estimated by (\ref{eq:est1}) and (\ref{eq:est2}) for $\lambda_0^{-1}\ge 1000$ and $\lambda_0^{-1}>2000$, which corresponds to $\mbox{ARL} \ge\lambda^{-1}\delta$. 
    
    To obtain the simulated ARL, we generate events in time window $[0,60000]$ and compute the run length when the statistics exceed the corresponding threshold. Note that this approximation will always underestimate the ARL since we can only generate events in a finite time window. We can see the thresholds computed from (\ref{eq:est1}) give us desired results. However, (\ref{eq:est2}) tends to overestimate the threshold. 
    
     \begin{table}[h]
        \begin{center}
        \caption{Verification of approximated ARL in (\ref{eq:est1}) and (\ref{eq:est2})}
        \begin{tabular}{|l|c|c|c|}
        \hline
             &  $b$ & theoretic ARL &simulated ARL\\\hline
            Results of (\ref{eq:est1}), $m=100$ & 3.3718 &10000& 9189\\\hline
            Results of (\ref{eq:est1}), $m=50$&3.3859&10000& 9561\\\hline
            Results of (\ref{eq:est2})&3.6625&10000& 21773\\\hline
            Results of (\ref{eq:est1}), $m=100$ & 3.5824 &20000& 17158\\\hline
            Results of (\ref{eq:est1}), $m=50$&3.5867&20000& 17655\\\hline
            Results of (\ref{eq:est2})&3.8352&20000& 41701\\\hline
        \end{tabular}
        \label{tbl:ARL}
        \end{center}
    \end{table}

    Now, let's compare the expected detection delay (EDD) of our proposed method with the generalized likelihood ratio (GLR) method in \cite{li2017detecting}. In the experiments of EDD, the distribution under $H_0$ is set as mentioned above. The thresholds of our methods are set according to the estimate of eq.(\ref{eq:est1}) with $m=50$, so that our desired ARL are 10000 or 20000 (see details in Table \ref{tbl:ARL}). As for the GLR, we compute the log generalized likelihood ratio with frequency $0.1$ per time unit and window length $w = 200$ with and without the cluster structure. For the GLR with the cluster structure (GLR-C), similar to the proposed method, we compute the statistic on each of the clusters and take the maximum. For the vanilla GLR we consider a change on the 16 edges in the union of the 4 clusters. The maximum likelihood estimates of the $\mathbf A_1$ and $\bm\mu_1$ are computed by the EM method. The thresholds of the desired ARLs are estimated with simulation. 
    
    We compare the performance of our methods with GLR and GLR-C in different settings and the results are shown in Table \ref{tbl:H1setting}. The EDDs are shown in columns 4-9 of Table \ref{tbl:EDD}. The results show that our proposed method achieves better performance when the change is within cluster and balanced on the edges (Case ($i$)-($iii$)). In Case $(iv),(v)$ where the change happens on multiple clusters and in Case $(vi)$ where the change happens on only part of a cluster, the proposed method is comparable to GLR with or without the cluster structure. Table \ref{tab:runtime} shows the advantage of our method: the computation time of our proposed methods is much less than the GLR methods, since our method does not require estimating the post-change distribution parameters. However in Case $(vii)$ if the change happens on a single edge, in other words, the clusters cannot accurately capture the topology of the local change, the performance of score based method can be worse than GLR.
    \begin{table}[h]
        \begin{center}        
        \caption{Setting of different cases in Table \ref{tbl:EDD}}
        \begin{tabular}{|l|c|}
        \hline
           &Changed parameters in post-change distribution\\\hline
           Case \textit{i}&$\alpha_{4,1} = \alpha_{4,3}= \alpha_{4,5}=\alpha_{4,8} = 0.2$\\\hline
           Case \textit{ii}& $\alpha_{4,1} = \alpha_{4,3}= \alpha_{4,5}=\alpha_{4,8} = 0.5$\\\hline
           Case \textit{iii}&  $\alpha_{4,1} = 0.6,\alpha_{4,3} = 0.4,\alpha_{4,5} = \alpha_{4,8} = 0.5$ \\\hline
           Case \textit{iv}&$\alpha_{4,1} = \alpha_{4,3} = \alpha_{9,5} = \alpha_{9,8} = 0.5$\\\hline
           Case \textit{v}&$\alpha_{4,5}=\alpha_{4,8} = \alpha_{9,8} = \alpha_{9,5} = 0.5$\\\hline
           Case \textit{vi}&$\alpha_{4,5} = \alpha_{4,8} = 0.5$\\\hline
           Case \textit{vii}& $\alpha_{4,5} = 0.5$\\\hline
        \end{tabular}
        \label{tbl:H1setting}
        \end{center}
    \end{table}
    
    \begin{table}[h]
        \begin{center}
        \caption{Comparison of EDD}
        \begin{tabular}{|c|c|c||c|c|c|c|c|c|c|}
        \hline
           Method&threshold&ARL&Case \textit{i} & Case \textit{ii} & Case \textit{iii} & Case \textit{iv} & Case \textit{v}&Case \textit{vi}&Case \textit{vii} \\\hline\hline
           Proposed&3.400&10000& \textbf{104.5} & \textbf{44.43} &\textbf{46.89} & 54.02 & 45.34 & 81.92 & 159.0 \\\hline
           GLR-C& 7.705&10000&108.9 & 48.77 & 47.77 &\textbf{51.38}&  \textbf{39.61} &  \textbf{68.71}&   \textbf{97.91}\\\hline
           GLR& 11.35 & 10000& 119.1  & 52.36 &  51.73 &  52.20&  41.06 &  74.23  & 105.6 \\\hline\hline
           Proposed&3.635&20000&\textbf{111.9} & \textbf{47.40} & \textbf{49.54} & 57.82 & 49.31 & 89.16 &  176.9 \\\hline
           GLR-C& 8.600&20000& 117.0 & 51.82 &  50.33 &  \textbf{55.74}&  \textbf{42.42} &  \textbf{72.85} &  \textbf{103.84} \\\hline
           GLR& 12.45 &20000 & 128.2 & 55.30 &  55.16  & 55.85&  42.81 &  79.49 &   112.5 \\\hline
        \end{tabular}
        \label{tbl:EDD}
        \end{center}
    \end{table}

    \begin{table}[htbp]
        \centering
        \caption{Comparison of computation time (in seconds) for computing the detection statistic over 50000 unit time.}
        \begin{tabular}{|c||c|c|c|}\hline
        Method & Proposed & GLR-C & GLR\\\hline\hline
          Duration (s)   & 3.789 & 20.47  & 72.97 \\\hline
        \end{tabular}
        
        \label{tab:runtime}
    \end{table}

\begin{figure}
        \centering
        \includegraphics[width=0.49\textwidth]{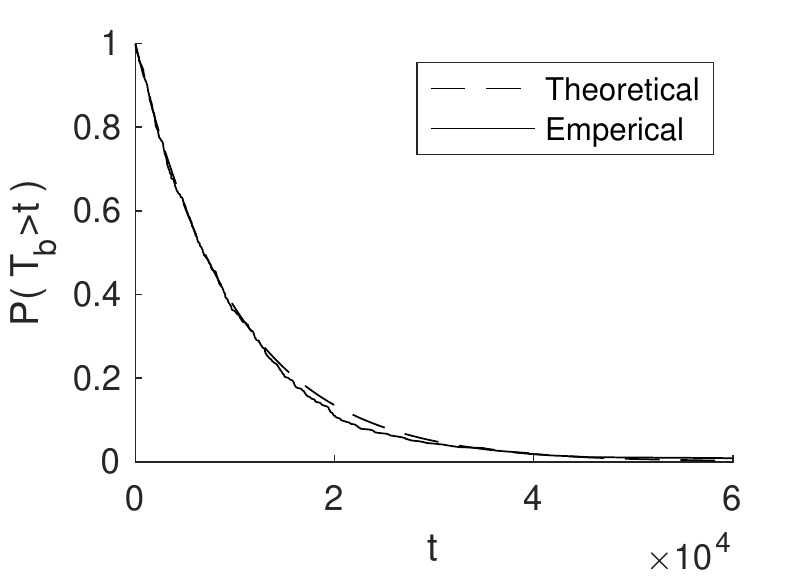}
        \includegraphics[width=0.49\textwidth]{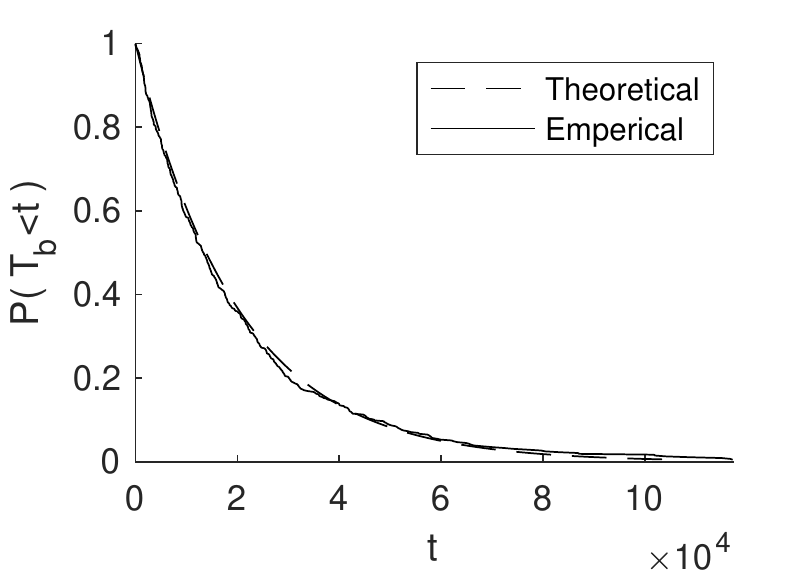}
        \caption{The probability $\mathbb P(T_b<t)$ under $H_0$ at $\text{ARL} = 10000$ and $\text{ARL} = 20000$.}
        \label{fig:exp_arl}
    \end{figure}

    Finally, we can verify the exponentiality of the run length by comparing the empirical c.d.f. over 500 experiments vs. the theoretical one, as shown in Figure \ref{fig:exp_arl}.

    \subsection{Real-data}
    In this section, we apply our scan statistics on a {\it memetracker data} and {\it stock data}. 
    \begin{itemize}
        \item memetracker data: It tracks texts and phrases, which are called memes, over different websites. This data is used to study information diffusion via social media and blogs. We use three meme data in \cite{li2017detecting}. The first data is ``Barack Obama was elected as the 44th president of the United States''. We use data from the top 40 news websites, which include Yahoo, CNN, Nydaily, The Guardian, etc. We use the data from Nov.01.2008 to Nov.02.2008 as the training data and the data from Nov.03.2008 to Nov.05.2008 as the test data. Our procedure detects a change at the time around 7 pm on Nov. 03, which is a few hours before the votes. 
        The second data is ``the summer Olympics game in Beijing''. We use data from Aug.01.2008 to Aug.03.2008 as the training data and data from Aug.04.2008 to Aug.15.2008 as the test set. 
        \item Stock price data: This data is downloaded from Yahoo Finance. We collect the closing price and trading volume of stock tickers: SPY, QQQ, DIA, EFA, and IWM, which are all index-type stocks and can reflect the situation of the overall stock market. For each ticker, we construct 3 types of events. High return: the day with a return over 90 percentile. Low return: the day with a return below 10 percentile. High volume: the day with trading volume over 90 percentile. Therefore, in this data, we have a network with 15 nodes. Such extreme trading events are of interest in the study of finance \cite{embrechts2011multivariate}. We use the data from Jan.04.2016 to Dec.31.2018 as the training data and data from Jan.01.2019 to Dec.31.2020 as the test data.
    \end{itemize}
    For each data, we apply the Newton method to fit the MLE of the parameters for the training set. Then, we use the fitted parameters to compute the scan statistics on the test set. For memetracker data, we construct the cluster by applying community detection methods on the fitted $\hat A$. For stock data, each cluster is the events that are related to a certain ticker. Details are shown in table \ref{tab:realData}. The change point detected from the ``Obama'' data is around 7 am on 2008.11.03. The result indicates that the public opinion of Barack Obama changed around one day before the votes. For the ``Olympic'' data, our procedure detects a change on Aug.04, three days before the Olympic game. For the stock data, we detect three-time intervals of which the starting dates are 2019.06.21, 2019.08.16, and 2020.03.04. According to the news, the first change-point 2019.06.21 is the date that the S\&P 500 hit a new record high, and the three major stock indexes surged on different scales. The second change-point 2019.08.16 is related to the US-China trade war. In August 2019, both US and China made multiple announcements about their tariffs. The last change-point is 2020.03.04, which is three business days before the first circuit breaker in 2020. There are many change-points after the first circuit breaker 2020.03.09, which indicates a long-term change in the stock market caused by pandemics and trade war. Real data results show that our proposed scan statistics achieve good performance in detecting the real change in different areas such as social media and financial markets.
    
    \begin{table}[H]
        \begin{center}
        \caption{result of real data}\resizebox{1\columnwidth}{!}{
        \begin{tabular}{|c|c|c|c|c|c|}
        \hline
             Data& training set & test set &\# of cluster & thresholds & change-points  \\\hline
             ``Obama''&08.11.01-08.11.02& 08.11.03-08.11.05&5&4& 11.03 7am\\
             ``Olympic''&08.08.01-08.08.03& 08.08.04-08.08.08&4&4& 08.04 6pm\\
             stock data&16.01.04-18.12.31&19.01.01-20.12.31&5& 4&19.06.21, 19.08.16, 20.03.04\\\hline
        \end{tabular}}
        \label{tab:realData}
        \end{center}
    \end{table}
    \begin{figure}[H]
        \centering
        \includegraphics[width=0.48\linewidth]{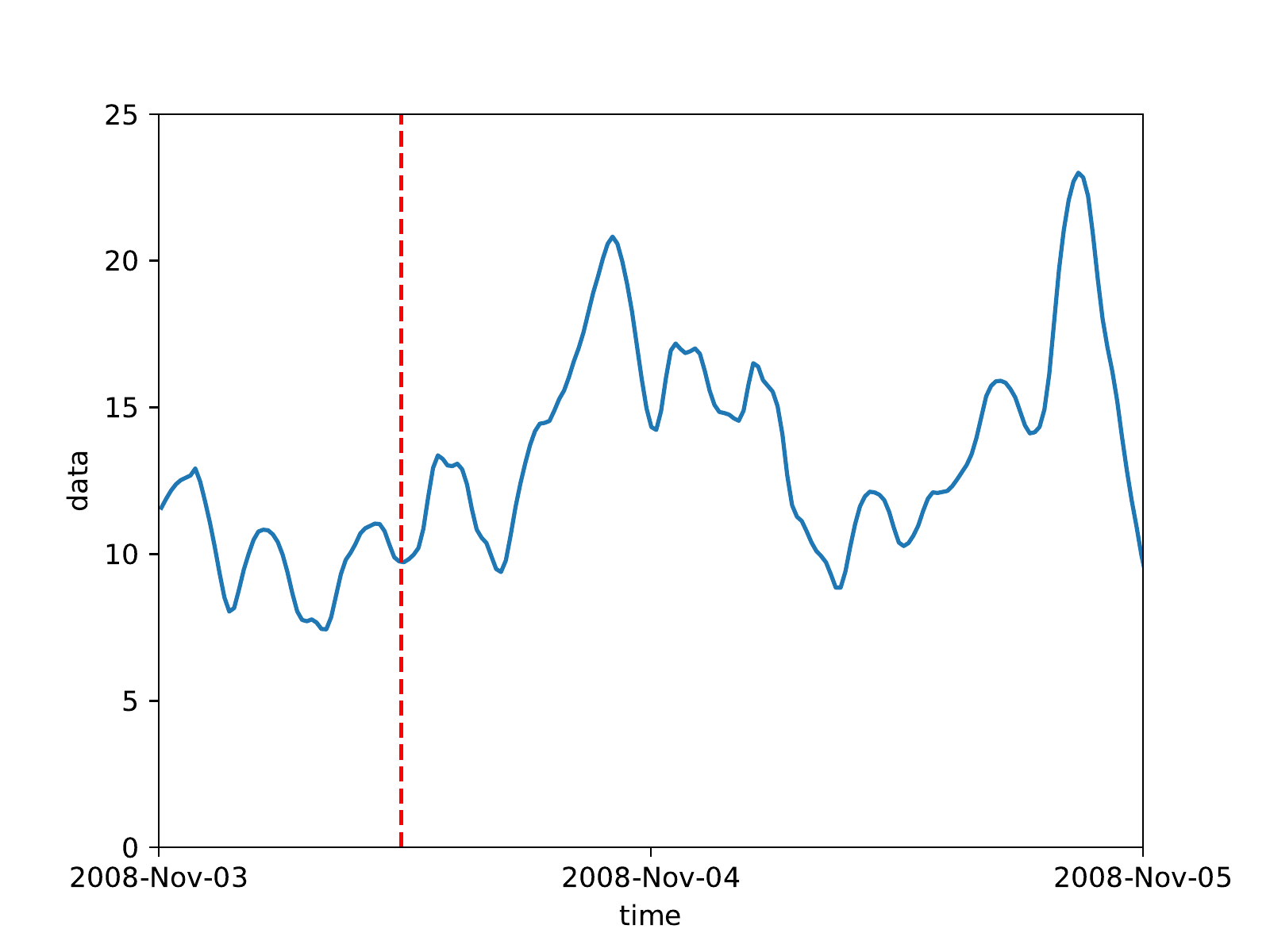}
        \includegraphics[width=0.48\linewidth]{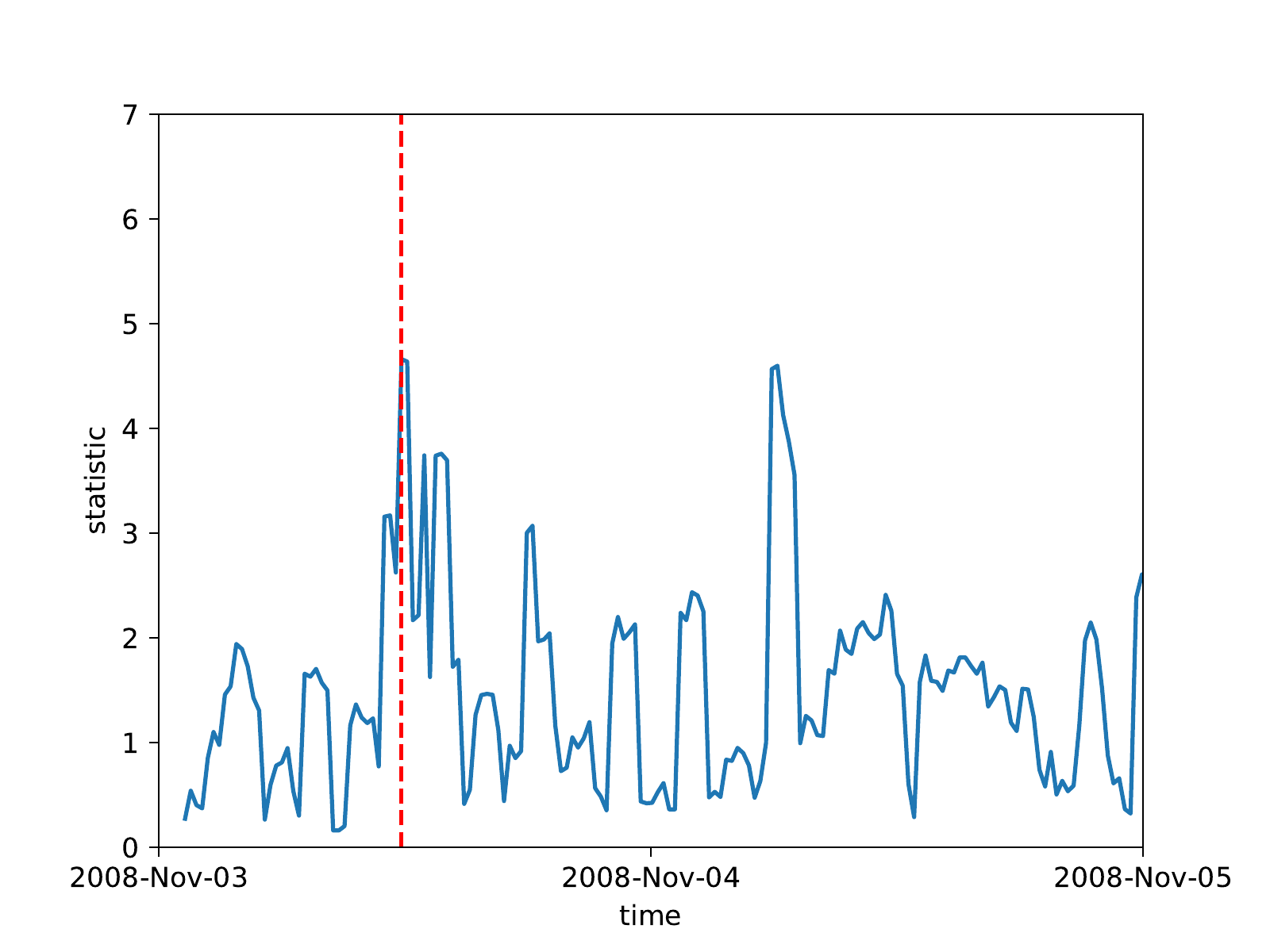}
        \includegraphics[width=0.48\linewidth]{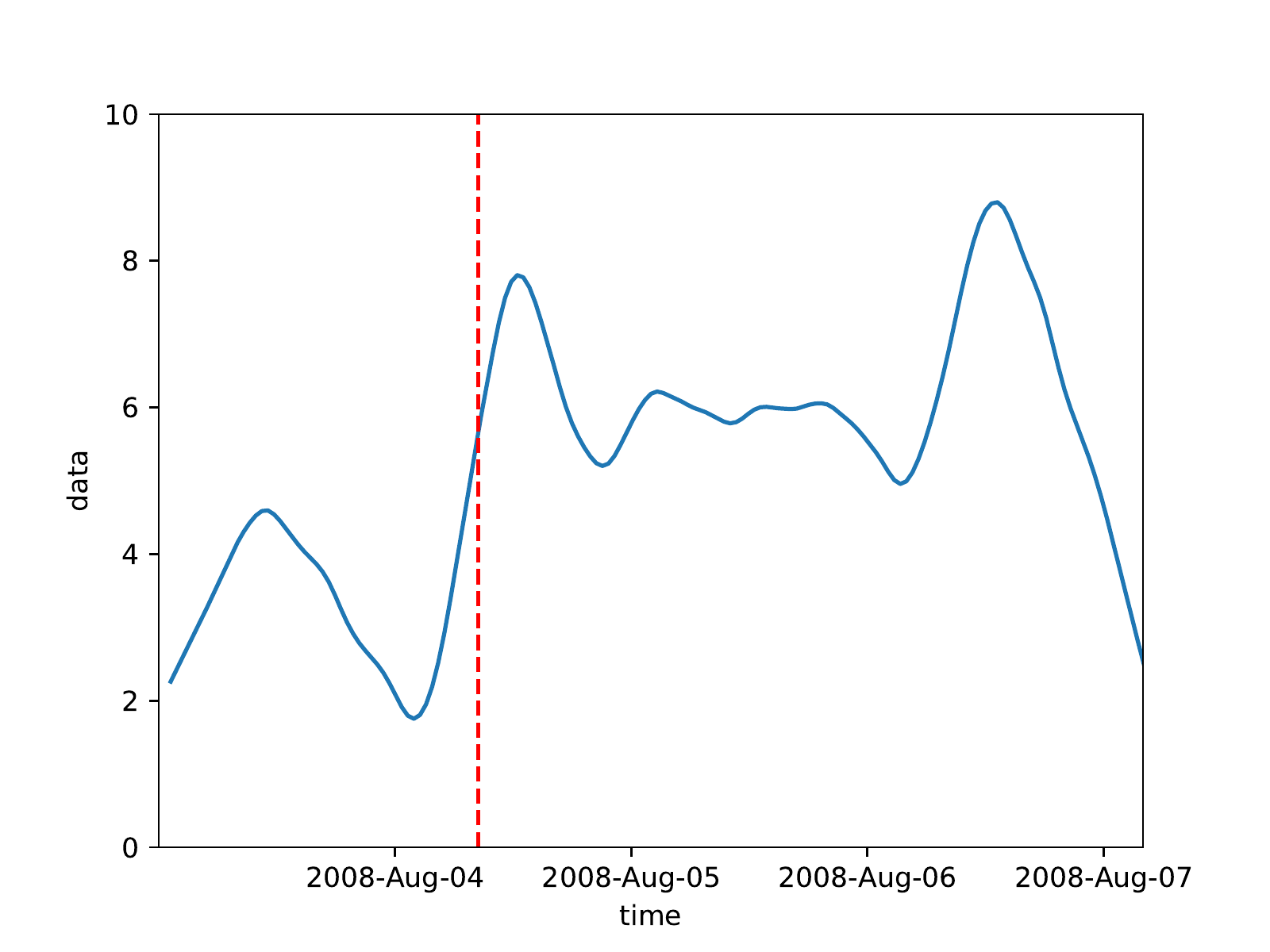}
        \includegraphics[width=0.48\linewidth]{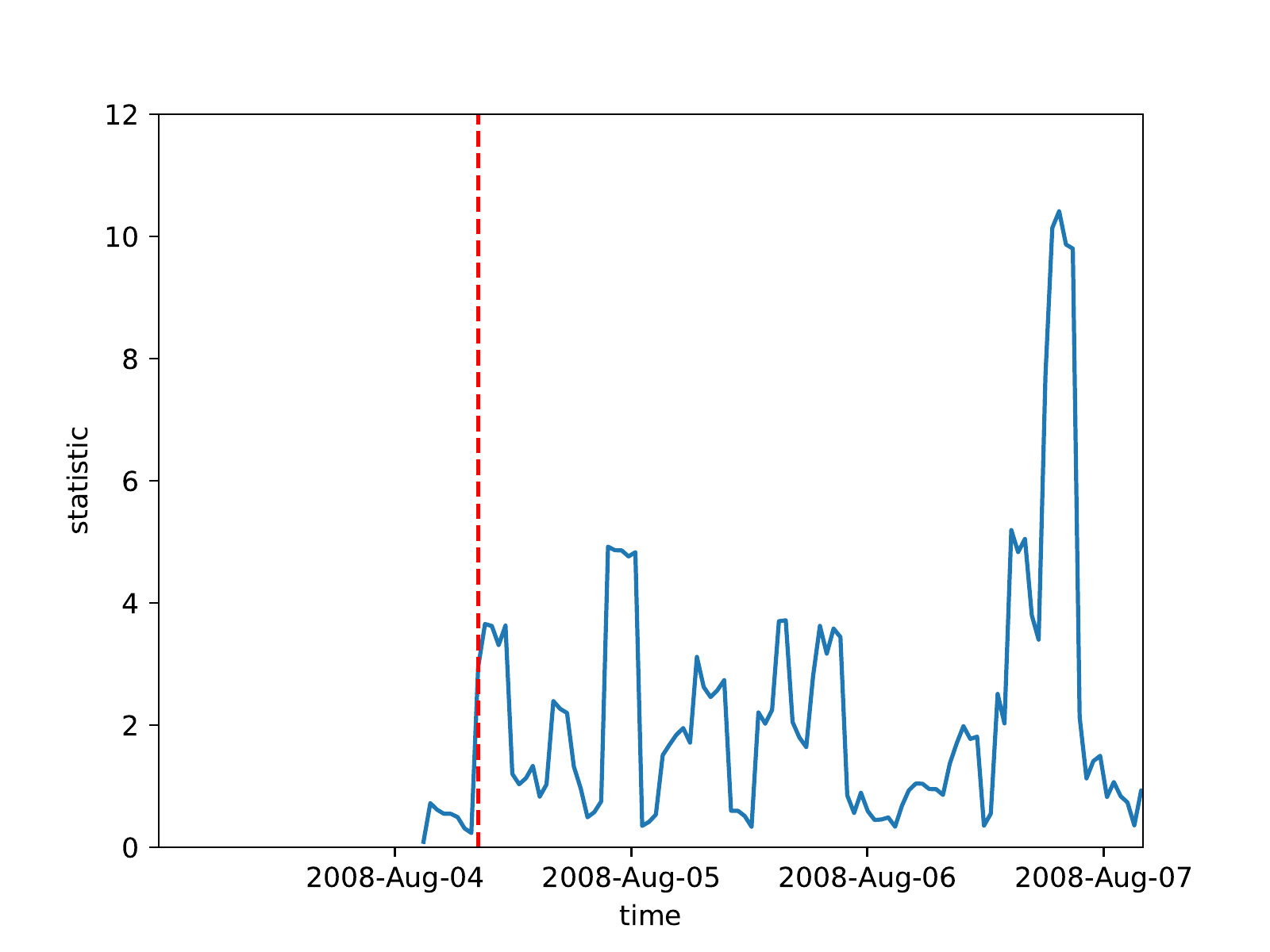}
        \includegraphics[width=0.48\linewidth]{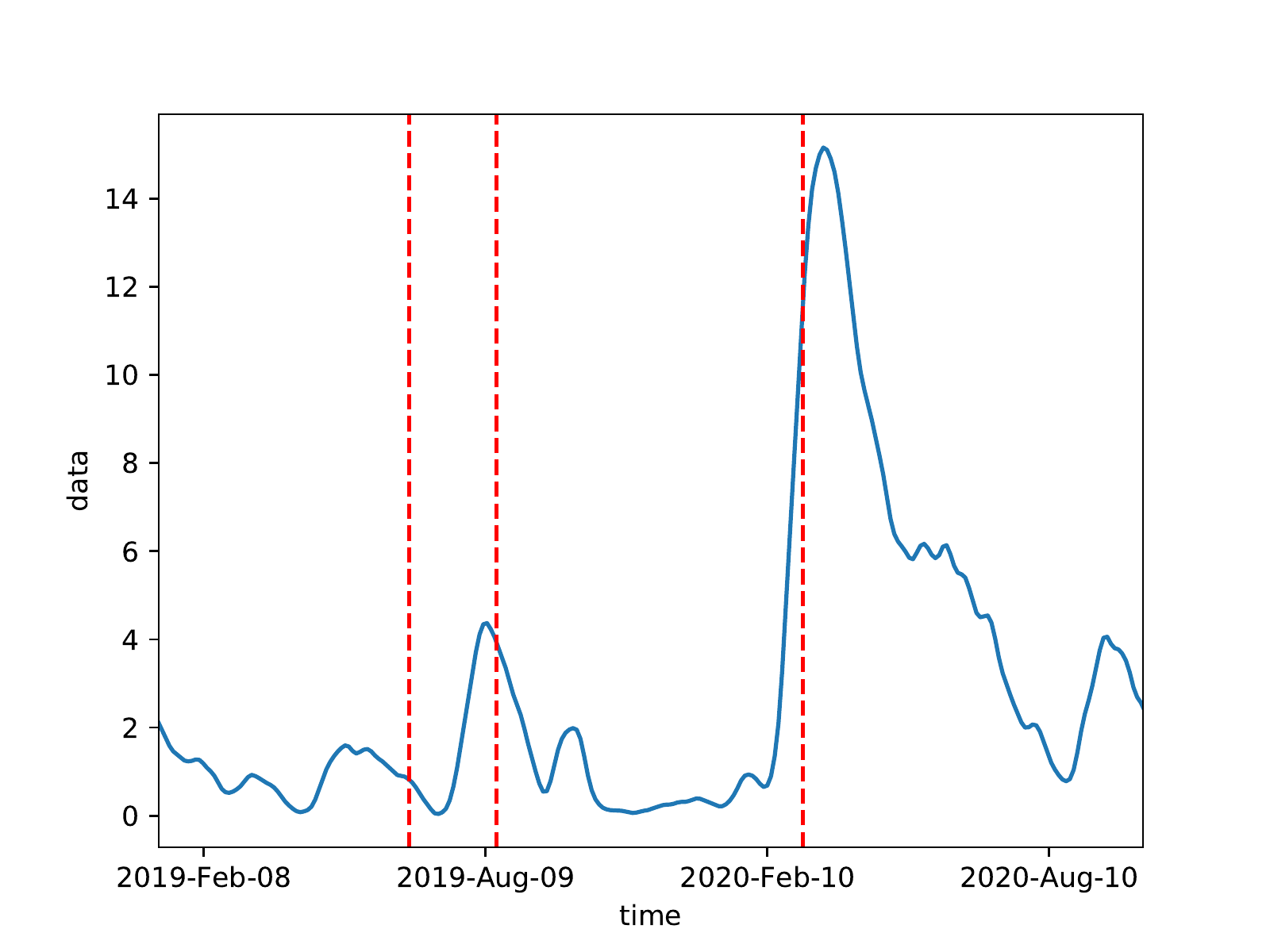}
        \includegraphics[width=0.48\linewidth]{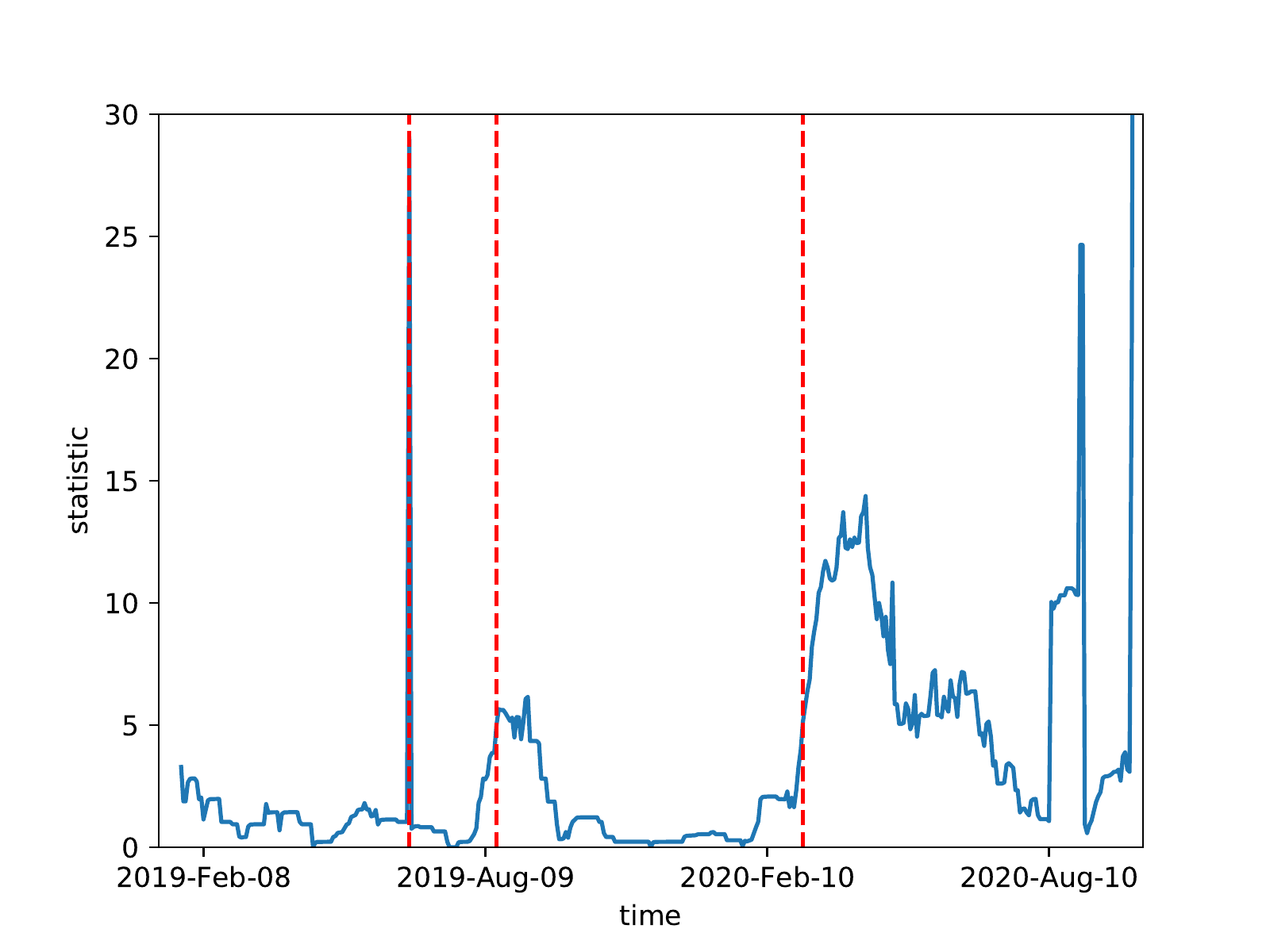}
        \caption{Scan Statistics procedure applied on real data.  For the upper plots, the blue line is the smoothed frequency of all events in the network. For the lower plots, the blue line is the scan statistics of proposed procedure. (1,1) \& (1,2): data of ``Obama''. (2,1) \& (2,2): data of ``Olympic''. (3,1) \& (3,2): Stock price data. Red line: detected change-points.}
        \label{fig:realData}
    \end{figure}

\section{CONCLUSION}

In this paper, we propose scan score statistics for detecting the change-points of network point processes. We use the multivariate Hawkes process to model the sequential event data. Our proposed method is based on score statistics without the requirement to estimate the post-change network parameters, which can be difficult to perform given limited post-change samples since we would like to detect the change quickly. In this sense, our method is more computationally efficient than the conventional GLR method, which is essential in online detection. We derive the asymptotic properties of the scan statistic, which enables us to provide further analysis of the instantaneous false alarm probability, false alarm rate, average run length and develop a computationally efficient procedure to calibrate the threshold for false alarm control. In experiments, we first use simulated data to verify our theoretical results. We also perform our method in real-world data, which shows a promising detection performance. Future work includes providing a more detailed discussion on the false discovery rate of localizing the unknown change (some initial results are in \ref{sec-fdr}).

\section*{ACKNOWLEDGEMENT}
The authors would like to thank the anonymous referees, an Associate Editor and the Editor for their constructive comments that improved the quality of this paper. The work is partially supported by NSF CCF-1650913, NSF DMS-1938106, NSF DMS-1830210, and CMMI-2015787.

\clearpage
\appendix
\renewcommand{\thesection}{Appendix \Alph{section}}
\renewcommand{\theequation}{\Alph{section}.\arabic{equation}}

 \section{Extension: False Discovery Rate of Change Localization}\label{sec-fdr}
     
     This section discusses an extension of the change-point detection: false identification after the change detection. After a change has been detected, it is sometimes also of interest to localize it and find the cluster where it happens. This corresponds to multiple hypothesis tests given all the information up to $t$. More specifically, at a given $t$, we check whether $\{|\Gamma_{t,w}^{(i)}|>b\}$ is true or not, for $i=1,\dots, L$. Our procedure stops, whenever there is at least one discovery, i.e. $\exists i$, such that $\{|\Gamma_{t,w}^{(i)}|>b\}$. Let $\kappa$ denote the number of such discoveries. Among these $\kappa$ discoveries, there are true discoveries and false discoveries. Let $V$ denote the number of false discoveries. Then the false discovery rate is defined as:
    \begin{equation}
        {\rm FDR} = \bbE(V/\kappa;\kappa>0),
    \end{equation}
    which is of interest in the study of scanning statistics. 
    
    \cite{siegmund2011False} provides an estimator for the FDR, under the assumptions: (i) $V$ is Poisson distributed with expected value $\rho$; and (ii) the number of false discoveries $V$ is independent of the number of true discoveries $\kappa - V$. The estimator is given by:
    \begin{equation*}
        \widehat{\rm FDR} = \rho/(\kappa+1).
    \end{equation*}
    In our procedure, assume that $L$ and $b$ is large, with similar proof of Theorem \ref{thm3}, we can also show that the first assumption is satisfied, i.e:
    \begin{equation}\label{eq-fdr}
        \rho \approx L\mathbb P\{|\Gamma_{t,w}^{(i)}|>b\}.
    \end{equation} As for the second assumption, similar to the discussion in \cite{siegmund2011False}, if each cluster does not largely overlap with each other, most of them are independent. In such a case, the false positives should be an approximately uniform distribution over all clusters. If the true signals do not frequently occur, a false positive is close to a true signal with a very small probability. Therefore, the second assumption is approximately satisfied too. 
    To control the false discovery rate, we only need to compute the threshold $b$ according to eq.(\ref{eq-fdr}) for a desired $\rho$. Recall that $\Gamma_{t,w}^{(i)}\sim \mathcal N(0,1)$.

    
    
\begin{figure}
        \centering
        \includegraphics{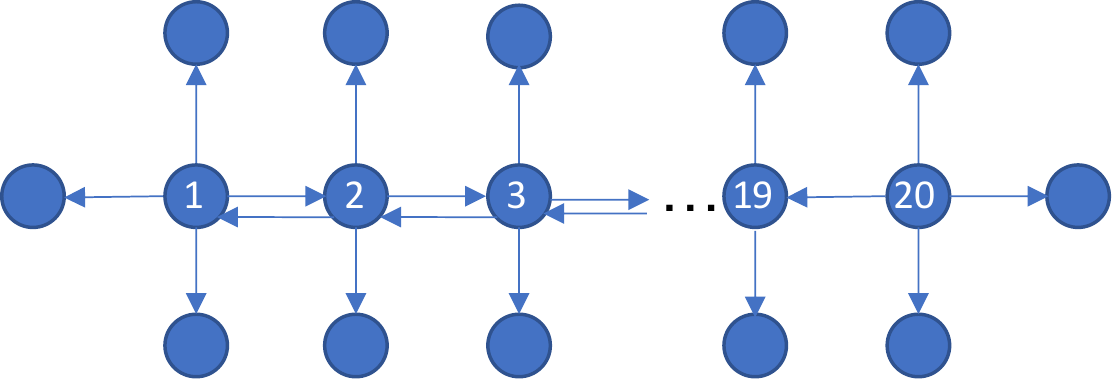}
        \caption{A network with 20 clusters in a line.}
        \label{fig:20_cluster}
    \end{figure}

    Below, we present several simulated examples to demonstrate that the simulated result of the estimation in eq.(\ref{eq-fdr}) provides an accurate approximation of the actual discovery rate. In this experiment, the network contains 20 clusters as shown in Figure \ref{fig:20_cluster}. All the clusters are placed in a line. Same as the previous simulated study, the pre-change distribution is a Poisson process with $\mu=1$. After $T_0$, the distribution of one cluster change to a Hawkes process with all the cross-excitation from the center to the neighboring nodes equal to 0.2. We generated the data of total length equal to 400 with $T_0=350$ and $T_1 = 50$. All the clusters are scanned with all the data up to 400. The experiments are repeated 200 times to compute the average number of false discovery, true discovery and false discovery rates. Table \ref{tbl:FDR} is the result. Notice that $\mathbb E(V) = \rho$ and it is close to the numbers in the last column, which shows a good approximation of eq.(\ref{eq-fdr}).

    \vspace{.1in}
            \begin{table}[h]
        \begin{center}
        \caption{Simulated value of false discovery rate, $t=w = 400$, $T_0 = 350$, $T_1 = 50$. Notice that the average number of false discovery $\mathbb E(V)$ is close to the number of last column which is the theoretical result.}
        \begin{tabular}{|c|c|c|c|c|}
        \hline
        $b$& $\mathbb E (V/(\kappa +1))$&$\mathbb E(V)$ & $\mathbb E (\kappa - V)$& $20\mathbb P(|\Gamma_{t,w}^{(i)}|>b)$\\\hline
        1.6&0.530&2.11&0.285&2.19\\\hline
        1.8&0.406&1.35&0.24&1.44\\\hline
        2&0.281&0.8&0.185&0.91\\\hline
        2.2&0.174&0.455&0.15&0.556\\\hline
        2.4&0.113&0.28&0.12&0.328\\\hline
        2.6&0.064&0.155&0.095&0.186\\\hline
        2.8&0.038&0.085&0.075&0.102\\\hline
        3.0&0.026&0.055&0.06&0.054\\\hline
        3.2&0.018&0.035&0.045&0.027\\\hline
        3.4&0.010&0.002&0.035&0.013\\\hline
        \end{tabular}
        \label{tbl:FDR}
        \end{center}
    \end{table}

 \section{Proof of Theorem \ref{thm1}}
        \begin{proof} \text{ \\}
        \begin{enumerate}
            \item[(i)] Since under $H_0$, $S_T^{(q,q)}(0)$ has the same the distribution as the univariate case. 
            \begin{align*}
                        \Var_{H_0}(T^{-\frac{1}{2}}S_T(0)) =& T^{-1}\Var(S_T(0))\\
                                        =& T^{-1} \Big(\frac{T}{2\beta} + 
            \frac{4\mu T - 1}{4\beta^2} \\
                                         &+\frac{e^{-2\beta T}}{4\beta^2} 
            - \frac{3\mu}{2\beta^3} - \frac{\mu e^{-2\beta T}}{2\beta^3} 
            +\frac{2\mu e^{-\beta T}}{\beta^3}\Big)\\
                                        \rightarrow& \frac{1}{2\beta}+
                                        \frac{\mu}{\beta^2} \,\,\, {\rm as} 
                                        \,\,\,T\rightarrow\infty.
            \end{align*}
            \item[(ii)] To prove the variance of $S_T^{(p,q)}$, we use the fact that $\Var_{H_0}[S_T^{(p,q)}(\mathbf 0)] = -\bbE_{H_0} [\frac{\partial S_T^{(p,q)}(\mathbf 0)}{\partial \alpha_{p,q}}]$. Then
            \begin{align}
                &\bbE_{H_0}[-\frac{\partial S_T^{(p,q)}(\mathbf 0)}{\partial \alpha_{p,q}}] \\
                = & \bbE[\frac{1}{\mu_q^2} \sum_{k\in\calC (q,T)} (\sum_{i\in\calC(p,t_k)}e^{-\beta (t_k-t_i)})^2]\nonumber\\
                = & \bbE[\frac{1}{\mu_q^2} \bbE[\sum_{k\in\calC (q,T)} (\sum_{i\in\calC(p,t_k)}e^{-\beta (t_k-t_i)})^2|N_q, N_p]]\label{condOnNumber}\\
                = & \bbE[\frac{N_q}{\mu_q^2} \bbE[(\sum_{i=1}^{N_p}Z_i(t_k))^2| N_p, t_k]]\label{condOntk}\\
                = & \bbE[\frac{N_q}{\mu_q^2}\bbE[\sum_{i=1}^{N_p} Z_i^2(t_k) +\sum_{i\neq j}^{N_p} Z_i(t)Z_j(t)|N_p, t_k]]\nonumber\\
                = & \bbE\Big[\frac{N_q}{\mu_q^2}\Big(N_p \bbE [Z_i^2(t_k))|t_k] + N_p(N_p-1)\bbE_{i\neq j} [Z_i(t_k)Z_j(t_k)|t_k]\Big)\Big]\nonumber\\
                = & \bbE\Big[\frac{N_q}{\mu_q^2}\Big(\frac{N_p}{2\beta T}(1 - e^{-2\beta t_k}) + \frac{N_p(N_p -1)}{\beta^2 T^2} (1-e^{-\beta t_k})^2\Big)\Big]\label{compZi}\\
                = & \frac{T}{\mu_q} (\frac{1}{2\beta} + \frac{\mu_q}{\beta^2}) + o(T), \label{compNpNq}
            \end{align} 
            where $N_q$ and $N_p$ are the number of events in $[0,T]$ on nodes $q$ and $p$ respectively. In eq(\ref{condOnNumber}), we use the fact that for Poissson process, the arrival times 
            follow $i.i.d.$ uniform distribution when it is conditional on the number of arrivals. With this fact, in eq(\ref{condOntk}), we define
            \begin{align*}
                Z_i(t) = 
                \begin{cases}
                    0\, &{\rm if}\, t_i\ge t,\\
                    e^{-\beta(t - t_i)} &{\rm if}\, t_i< t.
                \end{cases}
            \end{align*}
            Since $t_i \stackrel{i.i.d}{\sim} {\rm unif}[0,T]$, then
            \begin{align*}
                \bbE Z_i(t) =& \frac{1}{T}\int_0^{t} e^{-\beta(t - u)} du = \frac{1}{\beta T}(1-e^{-\beta t})\\
                \bbE Z_i^2(t) =& \frac{1}{T}\int_0^{t} e^{-2\beta(t - u)} du = \frac{1}{2\beta T}(1-e^{-2\beta t}),
            \end{align*}
            which proves the eq(\ref{compZi}). Since $N_p$ and $N_q$ follow Poisson distribution with mean $T\mu_p $ and $T\mu_q $, respectively. 
            Therefore eq(\ref{compNpNq}) is prooved.
            \item[(iii)]  Follow the similar techniques in (ii), we can prove 
            \[\Cov_{H_0}[T^{-\frac{1}{2}}S_T^{(p,q)}(\mathbf 0), T^{-\frac{1}{2}}S_T^{(p',q)}(\mathbf 0)] \rightarrow \frac{\mu_p\mu_{p'}}{\mu_q\beta^2}.\]
        \end{enumerate}
    \end{proof} 
    
\section{Proof of Theorem \ref{thm2}}
\begin{proof}
    Follow the definition in \cite{rathbun1996asymptotic}, let's define the kernel function,
     \begin{equation*}
        g(\bfs_1,\bfs_2,t) = \bfs_2^\top A\bfs_1 e^{-\beta t},
     \end{equation*} where $\bfs_i\in\bbR^{M}.$
     Then, we can define the conditional intensity function:
    \begin{align*}
        \Lambda(\bfs,t) =& \mu(\bfs) + \int_0^t\int_X g(\bfs,\bfu, t-r) N(d\bfu\times dr)\\
                     =& \mu(\bfs) + \sum_{t_i<t} \bfu_i^\top A\bfs\cdot e^{-\beta(t-t_i)},
    \end{align*}
    where $\bfu = e_m$, if $u_i=m$, and $e_m$ is the vector that $m$th entry is 1 and other entries are 0.
    Further, define a measure with delta function:
    \begin{align*}
        v(x) =& \sum_{i=1}^M\delta_{e_i}(x)\\
        \delta_{e_i}(x) =&
        \begin{cases}
            1\,\,\, {\rm if}\,\,\, x=e_i\\
            0\,\,\, {\rm o.w.}
        \end{cases}
    \end{align*}
    We can write the likelihood function as the following:
    \begin{equation*}
    \ell_T(A)  = \int_0^T\int_X \log\Lambda(\bfs,t;A)N(d\bfs\times dt) 
        - \int^T_0\int_X \Lambda(\bfs, t;A) v(d\bfs)dt.
    \end{equation*}
    We can easily check this define the same multivariate Hawkes process in eq.(\ref{lambda}), (\ref{kernel}), (\ref{eq:llh}). Define the function $\Delta$ as:
    \begin{equation*}
        \Delta_{(i,j), (p,q)} \triangleq \frac{\dot{\Lambda}_{i,j}\dot{\Lambda}_{p,q}}
        {\Lambda},
    \end{equation*} where $\dot{\Lambda}_{i,j}$ is the partial derivative of $\Lambda$ with respect to $\alpha_{i,j}$. Therefore, by the result in eq.(4.7) of \cite{rathbun1996asymptotic}, we have:
    \begin{equation*}
        \frac{1}{T}\sum_{k=1}^K\frac{\Delta(\bfu_k,t_k)}{\Lambda(\bfu_k,t_k)} \rightarrow \calI(A).
    \end{equation*}
    By direct computation, we can have the result of eq.(\ref{eq:estI})
\end{proof}

\section{Proof of Theorem \ref{TheoremFAR}}
    \begin{proof}
    For every $n\geq w/\delta-1$, let $p_n = \mathbb P(T_b = (n+1)\delta|T_b > n\delta)$. Then $\mathrm{FAR} = \sup_{n}p_n$. For the smallest $n$, clearly there is $p_n = \mathbb P(\Gamma_{n\delta}>b)$ equals to the instantaneous false alarm probability. Also there is for every $n$,
    $$
    p_n = \frac{\mathbb P(T_b = (n+1)\delta)}{\mathbb P(T_b>n\delta)}\leq \frac{\mathbb P(T_b > (n+1)\delta-w, \Gamma_{(n+1)\delta}>b)}{\mathbb P(T_b>n\delta)}.
    $$
    Since $\{T_b>(n+1)\delta - w\}$ and $\{\Gamma_{(n+1)\delta>b}\}$ are independent, we have
    $$
    p_n\leq \frac{\mathbb P(T_b > (n+1)\delta-w)}{\mathbb P(T_b>n\delta)}\mathbb P(\Gamma_{(n+1)\delta}>b) = \frac{\mathbb P(\Gamma_t>b)}{\prod_{k=n+1 - \lceil w/\delta\rceil}^{n-1}(1-p_k)}.
    $$
    As $b\to\infty$, the instantaneous false alarm probability $\mathbb P(\Gamma_t>b)$ goes to 0, and for large enough $b$, there exists some $p^* \geq\mathbb P(\Gamma_t>b)$, $p^* = \mathbb P(\Gamma_t>b)(1+o(1))$, such that $ \mathbb P(\Gamma_t>b)(1-p^*)^{1-C}= p^*$. Then by induction we can see for every $n$,
    $$
    p_n\leq \frac{\mathbb P(\Gamma_t>b)}{\prod_{k=n+1 - \lceil w/\delta\rceil}^{n-1}(1-p_k)}\leq \frac{\mathbb P(\Gamma_t>b)}{(1-p^*)^C}= p^*.
    $$
    \end{proof}

\section{Proof of Theorem \ref{thm3}}
\begin{proof}
    According to the Theorem I in \cite{arratia1989two}, let's define the ``neighbor of dependence'' for index $j$, $J(j) = \{(j-1), j, j+1\}$, with simple modification for $j=1$ and $j=k$, for $m>w$, $X_j$ and $X_i$ are independent for $i\not\in J(j)$. Therefore the dependence of elements not in the neighbor vanished, i.e. $b_3$ and $b_3'$ equals to 0.
    \begin{eqnarray}
        b_1 &=&\sum_{j=1}^k \sum_{i\in J(j)}\mathbb P(X_j=1)\mathbb P(X_i=1)\nonumber \\
            &\leq&3k\mathbb P(X_1=1)^2,\\
        b_2&=&\sum_{j=1}^k \sum_{i\in J(j)\setminus j}\mathbb P(X_j=1, X_i = 1)\nonumber \\
            &\leq&2k\mathbb P(X_1=1, X_2 = 1).\nonumber
    \end{eqnarray}
    Note that the event $\{X_1 = 1\}$ is the union of $\{\max_{0<n\leq m-w/\delta}\Gamma_{n\delta}\geq b\}$ and $\{\max_{m-w/\delta<n\leq m}\Gamma_{n\delta}\geq b\}$, and the former is independent of $X_2$. The same decomposition can be done to $X_2$. Then $b_2$ can be upper bounded by
    \begin{eqnarray}
            b_2&\leq&2k\left(\mathbb P(X_1 = 1)\mathbb P(X_2 = 1) + \mathbb P\left(\max_{m-w/\delta<n\leq m}\Gamma_{n\delta}\geq b,\max_{m<n\leq m+w/\delta}\Gamma_{n\delta}\geq b\right)\right)\nonumber\\
            &\leq& 2k\mathbb P(X_1 = 1)\mathbb P(X_2 = 1) + 2k\mathbb P\left(\max_{m-w/\delta<n\leq m}\Gamma_{n\delta}\geq b\right).
    \end{eqnarray}
    
    With the inequality of the tail probability of normal distribution in \cite{feller1968Anin}, 
    \begin{eqnarray}
        \mathbb P(X_1 =1) = \mathbb P\Big\{\max_{\substack{0<n\leq m,\\ 1\leq i\leq L}} |\Gamma_{n\delta,w}^{(i)}|>b\Big\}\nonumber \leq 2mL\bar\Phi(b)\leq\frac{2mL}{b}e^{-\frac{b^2}{2}},
    \end{eqnarray}
    where $\bar\Phi$ is the tail probability of standard normal random variable.
    With the same computation, we can show
    \[\mathbb P\left(\max_{m-w/\delta<n\leq m}\Gamma_{n\delta}\geq b\right)\leq \frac{2wL/\delta}{b}e^{-\frac{b^2}{2}}\]
    Therefore with the Theorem 1 in \cite{arratia1989two}, 
    \begin{eqnarray}
        &&|\mathbb P(T_b>xf(b)\delta) - e^{-\mathbb EW}| \\
        &=& |\mathbb P(W=0) - e^{-\mathbb EW}|\\
        &<&b_1+b_2\\
        &\leq& \frac{12km^2L^2}{b^2e^{b^2}}+ \frac{8km^2L^2}{b^2e^{b^2}} + \frac{4kwL/\delta}{be^{b^2/2}}\\
        &=& \frac{12xmL^2}{be^{b^2/2}} +\frac{8xmL^2}{be^{b^2/2}} + \frac{4xwL/\delta}{m},
    \end{eqnarray}
    which becomes small when $b\rightarrow \infty$ and $w/\delta\ll m\ll f(b)$.
    \end{proof}

\end{document}